\documentclass{iau}
\usepackage{graphicx}
\usepackage{subfigure}

\title[JD 11.~~Simulations of a compact source for G2] %% give here short title %%
{Hydrodynamical simulations of a compact source scenario for G2 }

\author[Alessandro Ballone et al.]   %% give here short author list %%
{A. Ballone$^{1,2}$, M. Schartmann$^{1,2}$, A. Burkert$^{1,2,3}$, S. Gillessen$^{2}$, R. Genzel$^{2}$, T.K. Fritz$^{2}$, F. Eisenhauer$^{2}$, O. Pfuhl$^{2}$, T. Ott$^{2}$}

\affiliation{$^1$University Observatory Munich, Scheinerstra{\ss}e 1, D-81679 M{\"u}nchen, Germany \\ email: {\tt aballone@mpe.mpg.de} \\[\affilskip]
$^2$Max-Planck-Institute for Extraterrestrial Physics, Postfach 1312, Giessenbachstra{\ss}e, D-85741 Garching, Germany \\[\affilskip]
$^3$Max-Planck Fellow}

\pubyear{2013}
\volume{303}  %% insert here IAU Symposium No.
\pagerange{119--126}
\jname{The GC: Feeding and Feedback in a Normal Galactic Nucleus}
\editors{Editors}
\begin{document}

\maketitle

\begin{abstract}

The origin of the dense gas cloud ``G2'' discovered in the Galactic Center
(\cite[Gillessen et al. 2012]{Gillessen12}) is still a debated puzzle. G2 might be a diffuse cloud or the result
of an outflow from an invisible star embedded in it. We present here detailed simulations of the evolution of winds on G2’s orbit. We find that the
hydrodynamic interaction with the hot atmosphere present in the Galactic Center and
the extreme gravitational field of the supermassive black hole must
be taken in account when modeling such a source scenario. We also find that in this
scenario most of the Br$\gamma$ luminosity is expected to come from the highly filamentary densest shocked wind material.
G2's observational properties can be used to constrain the properties of the outflow and our best model has a mass outflow rate
of $\dot{M}\mathrm{_w=8.8\times 10^{-8} M_{\odot} \;yr^{-1}}$ and a wind velocity of $v\mathrm{_w = 50 \;km/s}$. These values are
compatible with those of a young TTauri star wind, as already suggested by \cite{Scoville13}.
%\keywords{Keyword1, keyword2, keyword3, etc.}
\end{abstract}

\firstsection % if your document starts with a section,
\section{Introduction}

In the last year, the discovery of the object G2 in our Galactic Center (\cite[Gillessen et al. 2012, 2013a,b]{Gillessen12, Gillessen13a, Gillessen13b}) has caught the attention of the astronomical community. The object has been detected in the L-band with the infrared imager NACO and with the integral field spectrograph SINFONI at the VLT in Brackett-$\gamma$, He I and Paschen-$\alpha$ line emission. The detection in these bands has also been confirmed by \cite{Eckart13} and \cite{Phifer13}, even if the nature of this object is still controversial.
With the help of observations from the last 10 yr, \cite[Gillessen et al. (2012) and Gillessen et al. (2013a,b)]{Gillessen12, Gillessen13a, Gillessen13b} derived the dynamical properties of the object, finding that G2 is orbiting around the supermassive black-hole on a very eccentric orbit ($e\mathrm{\approx0.98}$), with pericenter at 2400 Schwarzschild radii, which G2 is expected to reach in early 2014.

There are two main scenarios for the nature of G2. The first one is the so-called \textit{diffuse cloud scenario} for which G2 is a dense and compact clump: in this case the L-band emitting material is a warm dust component with temperature $T\mathrm{_{dust}\approx550 \; K}$, while the line emission comes from an ionized gas ($T\mathrm{_{gas}\approx10^4 \; K}$) component, with roughly constant Br$\gamma$ luminosity $L\mathrm{_{Br\gamma} \approx 2\times 10^{-3} L_{\odot}}$ between 2004 and 2013. \cite{Gillessen12} have derived, from the observed size and Br$\gamma$ luminosity, a density of about $\mathrm{\rho_c \approx 6.1\times10^{-19}\; g\; cm^{-3}}$, with a corresponding mass of $M\mathrm{_{G2} \approx 1.7 \times 10^{28}\; g \approx 3}$ Earth masses. Interestingly, the observed orbit of G2 roughly lies in the plane of the clockwise rotating disk of young and massive stars ranging from $\mathrm{0.04 \; pc}$ to $\mathrm{0.5 \; pc}$ around the central hot bubble \cite[(Genzel et al. 2003, Paumard et al. 2006, Alig et al. 2013)]{Genzel03,Paumard06, Alig13}, so G2 could be a compact gas cloud that formed as a result of stellar wind interactions \cite[(Cuadra et al. 2006, Gillessen et al. 2012, Burkert et al. 2012)]{Cuadra06,Gillessen12,Burkert12}. \cite{Burkert12}, \cite{Schartmann12} and \cite{Anninos12} have studied in high detail the evolution and fate of such an object with the properties of G2.

The second scenario for G2 is the so-called \textit{compact source scenario}, for which G2 is the outflow from a star in its center. \cite{Murray-Clay12} have shown that the observed properties of G2 can be explained by gas outflowing from a photoevaporating protoplanetary disk and being tidally stripped while reaching SgrA* (a similar scenario has also been proposed by \cite[Miralda-Escud\'{e} 2012]{Miralda-Escude12}).
\cite{Meyer12} investigated the possibility that a nova, being on a similar orbit, could produce an expanding shell, while \cite{Scoville13} suggested that the observed emission could come from the tip of an inner, thin and \textit{cold bow shock}, produced by the wind of a TTauri star plunging into SgrA*.

The \textit{compact source scenario} has been studied up to now only with simplified analytical approximations. The aim of our work is to study these effects for a large range of outflow parameters with the help of hydrodynamical simulations with the Eulerian code PLUTO \cite[(Mignone et al. 2012)]{Mignone12}.

\section{Simulations setup}
To simplify our model, we simulated a single spherical wind moving in a two dimensional uniform grid in cylindrical coordinates $(Z,R)$, This led us to simplify our problem, assuming a zero angular momentum orbit. This is not a severe restriction when the source is far enough from pericenter, since the observed orbit has a very high eccentricity $e\;\mathrm{\simeq0.98}$ \cite[(Gillessen et al. 2013b)]{Gillessen13b}. We simulated almost the entire domain of the orbit from G2's apocenter, at $Z=-1.64\times10^{17}\mathrm{\;cm}$ \cite[(from the orbital derivation of Gillessen et al. 2013a)]{Gillessen13a}, to very close to SgrA*, fixing the frame of reference on the SMBH. The wind outflows are modeled with a mechanical approach, i.e. fixing constant density and velocity in a very small circular input region, in order to reproduce the correct injection of mass on its outer boundary. The input region is following G2's observed orbit with time. An adiabatic index $\Gamma=1$ is assumed. This choice is based on the assumption that the shocked wind is so dense and cools so fast that it can be treated as being isothermal \cite[(Scoville \& Burkert, 2013)]{Scoville13}.
The hot atmosphere is modeled following the density and temperature distribution used by \cite{Schartmann12}, while the supermassive black hole's gravitational field is assumed to be a Newtonian point-source with mass $M_\mathrm{{BH}}  = 4.31 \times 10^6 \; \mathrm{M_{\odot}}$ \cite[(Gillessen et al. 2009)]{Gillessen09} at $Z,R=0$.

We refer to \cite{Ballone13} for more numerical details.

\begin{figure}
% \vspace*{-2.0 cm}
\begin{center}
\includegraphics[scale=0.12]{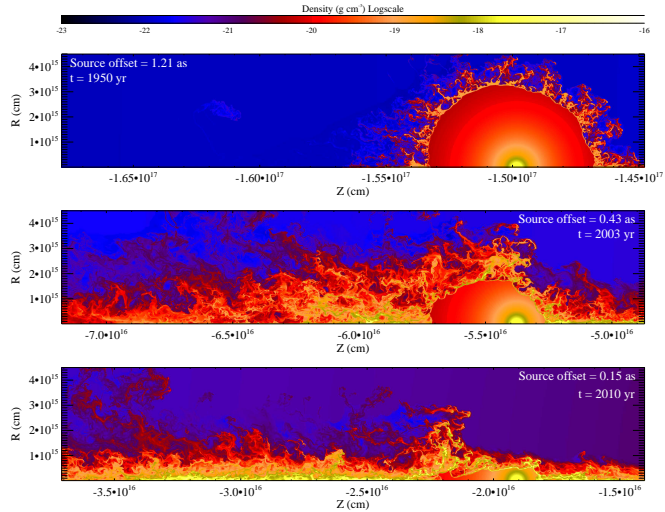} 
% \vspace*{-1.0 cm}

\label{fig1}
\end{center}
\caption{Density maps for our standard model, for source distances of $1''.21, 0''.43$, and $\mathrm{0''.15}$ from SgrA* (from top to bottom).}
\end{figure}

\section{Results and Discussion}

Our standard and best model has $\dot{M}\mathrm{_w=8.8\times 10^{-8} M_{\odot} \;yr^{-1}}$ and $v\mathrm{_w = 50 \;km\, s^{-1}}$. Fig. \ref{fig1} shows the evolution of the density with time (and with the motion towards the supermassive black hole).
Our simulations show that the presence of a surrounding high-temperature atmosphere (like that predicted by ADAF/RIAF solutions for the diffuse X-ray emission in the Galactic Center, e.g. \cite[Yuan et al. 2003]{Yuan03}) could be very important when modeling any \textit{compact source scenario} for G2. 
Due to the high pressure of the atmosphere, the structure of the studied winds is very different from that of typical stellar winds. As already shown by \cite{Scoville13}, the free-streaming wind interacting with this hot atmosphere will be shocked and already at early stages a very thin, dense and Rayleigh-Taylor unstable shell of shocked wind material forms around the free wind region. This is, along with the $1/r^2$ density distribution of the free-wind region, the main difference with respect to the \textit{diffuse cloud scenario}, where the object has instead a more or less uniform density all over its volume. Differently from the \textit{diffuse cloud scenario}, at late phases the ram pressure of the atmosphere can have an important role in effecting the structure of the wind (via stripping of wind material), while, as in the \textit{diffuse cloud scenario}, the dominant process at late phases is the squeezing and compression of the object in the direction perpendicular to the motion by the SMBH extreme tidal field. A simple coupling of all these different effects is hard to perform in an analytical study.

\begin{figure}\label{fig2}
 \begin{minipage}[b]{7.cm}
\begin{center}
\includegraphics[scale=0.38]{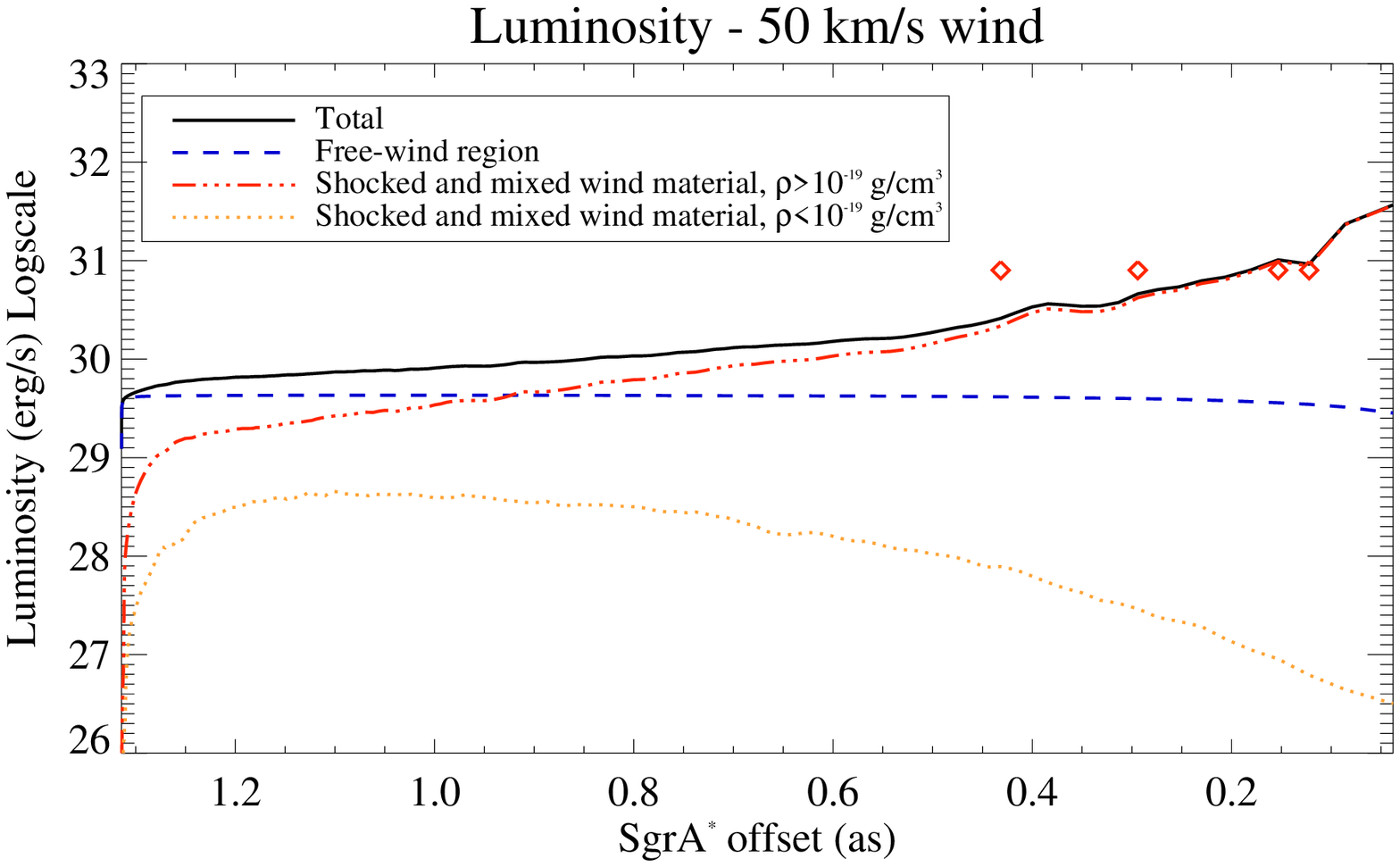} 
\end{center}
 \end{minipage}
 \ \hspace{-6.5mm} \hspace{3mm} \
 \begin{minipage}[b]{7.cm}
  \centering
\includegraphics[scale=0.19]{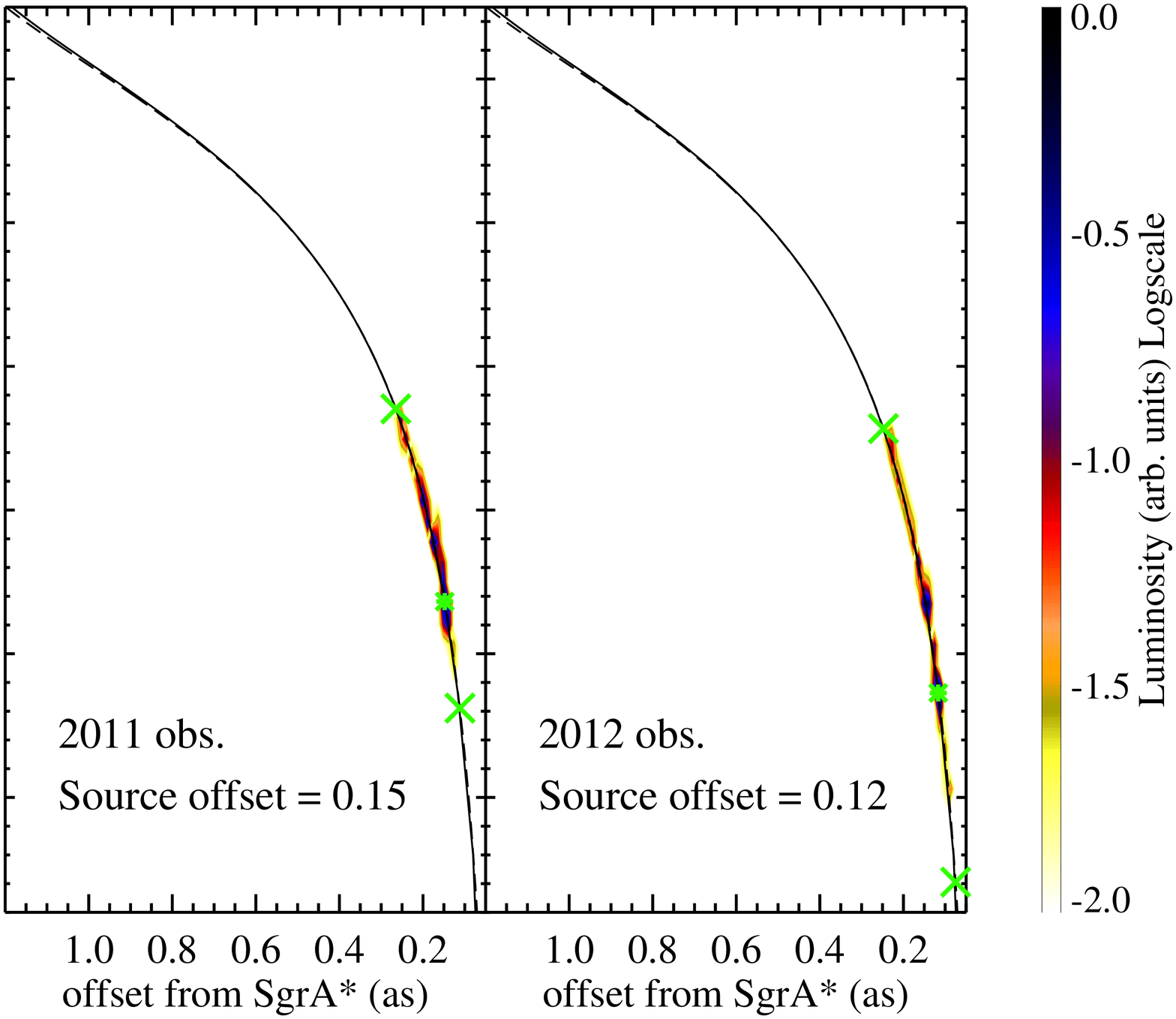}
 \end{minipage}
  \caption{Left panel: Br$\gamma$ luminosity evolution for our standard model. The solid line shows the total luminosity, the dashed line shows the luminosity of the free-wind region, the dash-dotted line shows the luminosity of the shocked wind with densities $>10^{-19}\;\mathrm{g\;cm^{-3}}$, and the dotted line shows the luminosity of the shocked wind material with densities $<10^{-19}\;\mathrm{g\;cm^{-3}}$. The red diamonds represent the observations.
Right: position-velocity diagrams for our standard model, for a source distance of $0''.15$, and $\mathrm{0''.12}$ from SgrA*. The green crosses show the G2 observed extremes and the green asterisk shows the position of the source in the diagram.}
\end{figure}

Fig. \ref{fig2} shows the evolution of the Br$\gamma$ luminosity. Our calculation for the best model gives a luminosity which is comparable with the observed ones, even if it increases significantly toward pericenter, while the observed one has roughly a constant value. Interestingly, most of the luminosity in such a scenario comes from the filamentary densest shocked wind material, as visible in the lower panel of Fig. \ref{fig2}. This is another main difference with the \textit{diffuse cloud scenario}, where the emission is generated by diffuse gas with rather uniform density. Fig. \ref{fig2} also shows that the size of G2 in the position-velocity diagrams can also be reproduced by the standard model.

\begin{figure}[!h]
% \vspace*{-2.0 cm}
\begin{center}
\includegraphics[trim= 2cm 4cm 2cm 4cm, clip=true, scale=0.24]{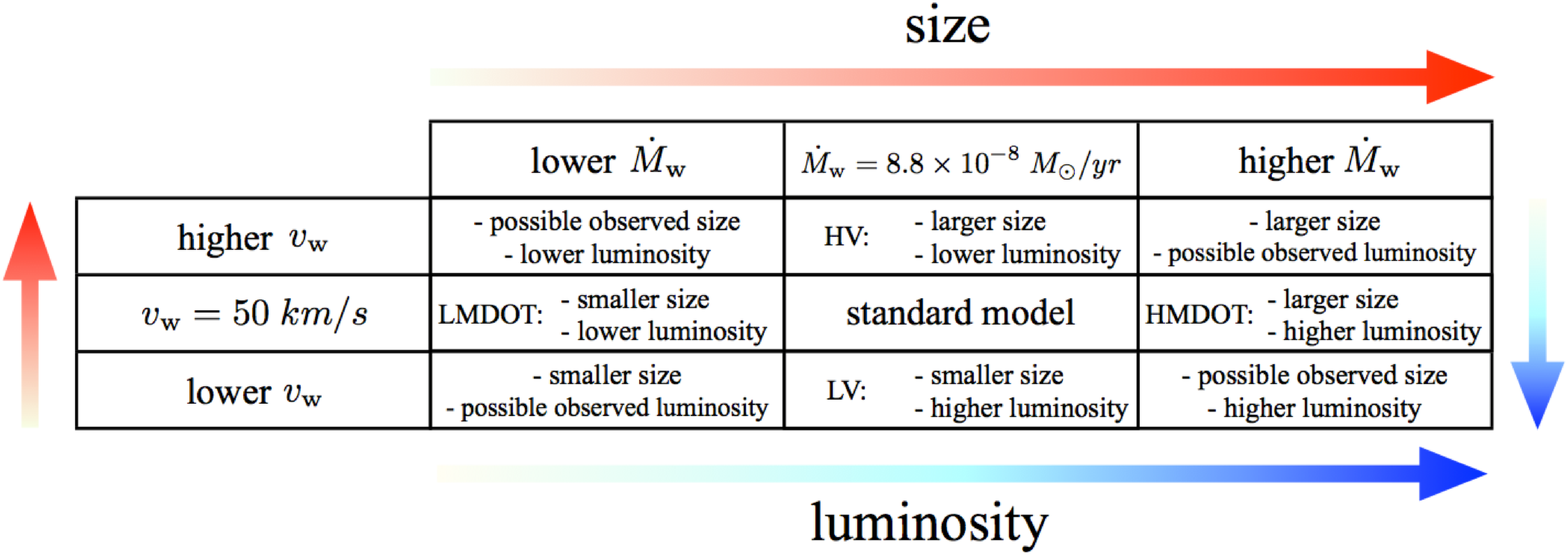} 
%\includegraphics[scale=0.17]{BalloneAfig3.eps} 
% \vspace*{-1.0 cm}
\caption{Dependence of G2 luminosity and size on the wind parameters.}
\label{fig3}
\end{center}
\end{figure}

We have finally made a small wind parameters study to constrain the properties of the outflow. We have hence divided and multiplied the wind mass-loss rate and velocity by a factor of 5, respectively. Such a small factor already gives significant differences in the resulting observational properties (e.g., roughly a factor of 10 different luminosities). As summarized in Fig. \ref{fig3}., when fixing the mass-loss rates, a higher velocity gives a lower luminosity and a larger size of the emitting material (and vice versa). At constant velocity, a higher mass loss rate is instead leading to a higher luminosity and a larger size (and vice versa). Thus, a combination of observed size and luminosity can effectively constrain the wind parameters. For our choice of the atmosphere, the wind parameters of our best model are comparable with those of a young TTauri star wind \cite[(White \& Hillenbrand 2004)]{White04}. The age of TTauri stars is also consistent with the age \cite[($\simeq\mathrm{6\pm 2 \;Myr}$, Paumard et al. 2006)]{Paumard06} of the clockwise disk of young stars, where the source could have been scattered from \cite[(Murray-Clay \& Loeb 2012)]{Murray-Clay12}.


\begin{thebibliography}{}

\bibitem[Alig et al. (2013)]{Alig13}
{Alig, C., Schartmann, M., Burkert, A. \& Dolag, K.} 2013, 
\textit{ApJ}, 771, 119

\bibitem[Anninos et al. (2012)]{Anninos12}
{Anninos, P., Fragile, P.C., Wilson, J. \& Murray, S.D.} 2012, 
\textit{ApJ}, 759, 132

\bibitem[Ballone et al. (2013)]{Ballone13}
{Ballone, A., Schartmann, M., Burkert, A., Gillessen, S., Genzel, R. et al.} 2013, 
\textit{ApJ}, 776, 13

\bibitem[Burkert et al. (2012)]{Burkert12}
{Burkert, A., Schartmann, M., Alig, C., Gillessen, S., Genzel et al.} 2012, 
\textit{ApJ}, 750, 58

\bibitem[Cuadra et al. (2006)]{Cuadra06}
{Cuadra, J., Nayakshin, S., Springel, V. \& Di Matteo, T.} 2006, 
\textit{MNRAS}, 366, 358

\bibitem[Eckart et al. (2013)]{Eckart13}
{Eckart, A.,  Mu\u{z}i\'{c}, K., Yazici, S., Sabha, N., Shahzamanian et al.} 2013, 
\textit{A\&A}, 551, A18

\bibitem[Genzel et al. (2003)]{Genzel03}
{Genzel, R., Sch\"{o}del, R., Ott, T., Eisenhauer, F., Hofmann, R. et al.} 2003, 
\textit{ApJ}, 594, 812

\bibitem[Gillessen et al. (2009)]{Gillessen09}
{Gillessen, S., Eisenhauer, F., Trippe, S., Alexander, T., Genzel, R. et al.} 2009, 
\textit{ApJ}, 692, 1075

\bibitem[Gillessen et al. (2012)]{Gillessen12}
{Gillessen, S., Genzel, R., Fritz, T.K., Quataert, E., Alig, C. et al.} 2012, 
\textit{Nature}, 481, 51

\bibitem[Gillessen et al. (2013a)]{Gillessen13a}
{Gillessen, S., Genzel, R., Fritz, T.K., Eisenhauer, F., Pfuhl, O. et al.} 2013a, 
\textit{ApJ}, 763, 78

\bibitem[Gillessen et al. (2013b)]{Gillessen13b}
{Gillessen, S., Genzel, R., Fritz, T.K., Eisenhauer, F., Pfuhl, O. et al.} 2013b, 
\textit{ApJ}, 774, 44

\bibitem[Meyer \& Meyer-Hofmeister (2012)]{Meyer12}
{Meyer, F., \& Meyer-Hofmeister, E.} 2012, 
 \textit{A\&A}, 546, L2

\bibitem[Mignone et al. (2012)]{Mignone12}
{Mignone, A., Zanni, C., Tzeferacos, P., van Straalen, B., Colella, P. et al.} 2012, 
 \textit{ApJS}, 198, 7

\bibitem[Miralda-Escud\'{e} (2012)]{Miralda-Escude12}
{Miralda-Escud\'{e}, J.} 2012, 
 \textit{ApJ}, 756, 86

\bibitem[Murray-Clay \& Loeb (2012)]{Murray-Clay12}
{Murray-Clay, R.A., \& Loeb, A.} 2012, 
 \textit{NatCo}, 3, 1049

\bibitem[Paumard et al. (2006)]{Paumard06}
{Paumard, T., Genzel, R., Martins, F., Nayakshin, S., Beloborodov et al. } 2006, 
\textit{ApJ}, 594, 812

\bibitem[Phifer et al. (2013)]{Phifer13}
{Phifer, K., Do, T., Meyer, L., Ghez, A.M., Witzel et al.} 2013, 
\textit{ApJL}, 773, L13

\bibitem[Schartmann et al. (2012)]{Schartmann12}
{Schartmann, M., Burkert, A., Alig, C., Gillessen, S., Genzel, R. et al.} 2012, 
\textit{ApJ}, 750, 58

\bibitem[Scoville \& Burkert (2013)]{Scoville13}
{Scoville, N., \& Burkert, A.} 2013, 
 \textit{ApJ}, 768, 108

\bibitem[White \& Hillenbrand (2004)]{White04}
{White, R.J., \& Hillenbrand, L.A.} 2004, 
 \textit{ApJ}, 616, 998

\bibitem[Yuan et al. (2003)]{Yuan03}
{Yuan, F., Quataert, E. \& Narayan, R.} 2003, 
 \textit{ApJ}, 598, 301

\end{thebibliography}
\end{document}